\documentclass[a4paper,10pt]{article}
\usepackage[utf8]{inputenc}

\usepackage{amsmath}
\usepackage{amssymb}
\usepackage{hyperref}
\usepackage{graphicx}
\usepackage{verbatim}
\usepackage{geometry}
\usepackage{tikz}
\tikzstyle{every picture}=[level distance = 8mm, baseline=-0.5ex]
\tikzstyle{prop}=[shape=circle,minimum size=6mm, draw=black!80, fill=green!30]

\newcommand{\acc}{\text{acc}}
\newcommand{\res}{\text{res}}

\newcommand{\C}{\mathbb{C}}
\newcommand{\R}{\mathbb{R}}

\begin{document}

\title{Analyticity domain of a Quantum Field Theory and Accelero-summation}
\author{Marc~P.~Bellon${}^{1}$, Pierre~J.~Clavier${}^{2}$\\
\normalsize \it ${}^1$ Sorbonne Université, CNRS, Laboratoire de Physique Théorique et Hautes
Energies,\\ \normalsize \it LPTHE, 75005 Paris, France\\
\normalsize \it $^2$ Potsdam Universit\"at, Mathematik, Golm, Deutschland }

\date{}

\maketitle

\begin{abstract} 
From 't Hooft's argument, one expects that the analyticity domain of an asymptotically
free quantum field theory is horned shaped. In the usual Borel summation, the function is obtained through 
a Laplace transform and thus has a much larger analyticity domain. However, if
the summation process goes through the
process called acceleration by \'Ecalle, one obtains such a horn shaped analyticity domain.
We therefore argue that acceleration, which allows to
go beyond standard Borel summation, must be an integral part of the toolkit for the study of
exactly renormalisable quantum field theories.
We sketch how this procedure is working and what are its consequences.
 
\end{abstract}

\textbf{Mathematics Subjects Classification:} 81Q40, 81T16, 40G10.


\section*{Introduction}

Recently Borel--Écalle resummation theory~\cite{Ecalle81,Ecalle81b,Ecalle81c} has seen a renewed interest in the physics community. 
It has been applied in the context of quantum field theories \cite{DuUn12,ChDoDuUn14}, the WKB approximation~\cite{DuUn14}, 
string theory~\cite{AnScVo12,CoEdScVo14}, matrix theory~\cite{MaScWe08} and the study of Schwinger-Dyson equation~\cite{BeCl16}.

The goal of this note is to detail some results announced in~\cite{BeCl16} relating accelero-summation to the analyticity domain of a quantum
field theory. In~\cite{Ho79}, through a clever use of renormalisation group invariance,
't Hooft argued that the two-point function of an asymptotically free theory cannot depend analytically on the coupling when it is in a disk
tangent to the origin, but only when it is in a horn-shaped domain. We show that this
domain is exactly the analyticity domain of a function obtained by a certain type of
accelero-summation.

This paper is organised as follows: in the first section, we recall this argument of 't Hooft.
Section~2 is devoted to a discussion of the link between the analyticity domain of a function and the way its expansion can be resummed. 
Section~3 gives a very short idea of what is acceleration. Finally, in section~4, 
we show how a particular kind of accelero-summation gives 't Hooft's domain of analyticity. The last section is devoted to the study of some consequences of this 
procedure.

\section{ 't Hooft's argument}

Consider an asymptotically free quantum field theory in the limit $\mu\to\infty$. The beta function of the theory is
\begin{equation*}
 \beta(a) =\mu\frac{d}{d\mu}a = -\beta_0a^2+\beta_1a^3 + \cdots
\end{equation*}
with $\beta_0>0$, for a perturbative parameter $a=g^2$ suitably scaled. 
Integrating the renormalisation group equation
\begin{equation*}
 \frac{d}{d\log\frac{p^2}{\mu^2}}\bar a=\beta(\bar a)
\end{equation*}
in the one loop approximation, that is neglecting the $a^3$ and higher terms in the beta function, one gets the classical result
\begin{equation} \label{renorm_fine_structure_cst}
 \bar a = \frac{a}{1+a\beta_0\log\frac{p^2}{\mu^2}}.
\end{equation}
The boundary value $\bar a(p^2=\mu^2)=a$ has been used. 

The observation of~\cite{Ho79} was that any two point function $G(a,{p}^2)$ of the theory has to be a function of the argument
\begin{equation}
 X = a^{-1}+\beta_0\log\frac{p^2}{\mu^2}.
\end{equation}
We are looking for the analyticity domain of this two point function in terms of the parameter $a$. We expect that 
when $a$ is real, the poles of $G$ are for $p^2+m^2=0$, with $m$ the physical mass of a particle. Therefore, for $a$ real, 
the only possible singularities of $G$ as a function of ${p}^2$ are when ${p}^2$ is real and strictly negative. Using the 
principal branch of the logarithm, we get that the singular points are for the following values of $X$ 
\begin{equation*}
 X_\text{sing}=r + \beta_0(2n+1)i\pi
\end{equation*}
for some $n$ in $\mathbb{Z}$ and $r=a^{-1} + \beta_0 \log(m^2/\mu^2)$ in $\mathbb{R}$. For any $p^2$ real and positive, we get the following complex
values of $a^{-1}$ for which $G$ has a pole 
\begin{equation*}
 a^{-1}_\text{sing} = r-\beta_0\log\frac{p^2}{\mu^2}+\beta_0(2n+1)i\pi.
\end{equation*}
Since a singularity in the euclidean domain is a sign of instability, we conclude that the theory is not defined when the imaginary part
of~\(a^{-1}\) is \( (2n+1) \beta_0 \pi\), which define lines \(L_n\).  

Let us recall once and for all the relation in conformal geometry between lines and circle by the inversion \(z \to z^{-1}\), 
since it will be used many times in this article to relate limits for the inverse coupling and those for the coupling itself.
We have
\begin{equation}
  \frac 1 z - \frac 1 {z+\bar z} = \frac {z + \bar z - z}{z(z+ \bar z)} = \frac {\bar z} z \frac 1 { z + \bar z}.
\end{equation}
Taking the norm of both sides and using that \(z\) and \(\bar z \) have the same norm, we obtain
\begin{equation*}
 \left|\frac{1}{z} - \frac{1}{2\Re(z)}\right| = \frac{1}{2\Re(z)}.
\end{equation*}
Thus we see that if \(z\) varies while its real part
\(\Re(z) = \frac 1 2 (z + \bar z)\) remains fixed, then \(z^{-1}\) remains at a constant distance of the fixed point \( (z + \bar z)^{-1}
\), so that it remains on a circle. Since the radius of this circle is equal to the distance of the centre to the origin, this circle
includes the origin.  The same relation between a line and a circle exists also when the line has an other direction, as can be seen by the
fact that a rotation of \(z\) around the origin by an angle \(\alpha\), given by the multiplication by \(\exp( i \alpha)\) results in a
rotation by an angle \(-\alpha\) of \(z^{-1}\).

In particular, by a multiplication by \(i\), one easily check that the line $L_n$ transforms in the circle of radius $|K_n|$ and of center $-iK_n$  with
\begin{equation*}
 K_n = \left(2\beta_0(2n+1)\pi\right)^{-1}.
\end{equation*}
Let us emphasize a few key points.

First observe that $|K_n|$ decreases when $|n|$ increases. The maximum values are
\begin{equation} \label{eq:limiting_circles}
 K_0 = \left(2\beta_0\pi\right)^{-1} = -K_{-1}.
\end{equation}
Second, all the singular values $a_n$ will lie inside the two disks $D_\pm$ of radius $K_0$ and of center 
$\pm iK_0$. 

Finally \(K_n\) goes to 0 as $n$ goes to infinity, therefore the poles will concentrate around the origin. Thus we find that the domain of analyticity of $G$ arbitrarily close to the 
origin is delimited by the circles $D_+$ and $D_-$ tangent to the origin. This is in stark
contradiction with the analyticity domain of 
Borel-summed functions. We thus have an argument that $G$ cannot be Borel summable, but we will see
that accelero-summation gives rise to such an analyticity domain.

\section{Summation methods and analyticity domains.}

There is a strong relation between the analyticity domain of an analytic function and the properties of
its expansion and it goes both way.  Indeed, in the simplest case of a convergent series, one obtains
an analytic function on the convergence disk of the series.  What is also true is that if one knows
that a function is analytic on a disk, Cauchy's theorem allows to recover the function from its value on the edge
of the disk, in a form which allows to prove that the function is the sum of its Taylor series at the
center of the disk.
Divergent power series therefore stem from the expansion of a function around a point at the
boundary of its domain of definition.

The computation of a function through classical Borel summation
stems from a Laplace transform.  The imaginary part of the variable \(z\) has no
effect on the modulus of \(\exp(-\xi z)\), so that the domain of convergence of this definition will
be a half space, when the expansion point is taken to be at infinity. {By this we mean that the 
domain of convergence, which is contained in the domain of analyticity, will be of the form 
\begin{equation*}
 \left\{z\in\C:\Re(z)>R\right\}
\end{equation*}
for some $R>0$.}
If we go back to a vicinity of
zero, this means that the Borel summation is defined on a disk with zero on its edge, as a consequence
of the relation between lines and circles seen in the preceding section.  It has been
possible to show that reciprocally, an analytic function defined in a half space neighborhood of
infinity has an expansion in inverse powers of \(z\) which is Borel summable, because the Laplace
transformation can be inverted by taking an integral on an infinite line parallel to the imaginary axis.  

For real physical quantities, the Laplace transform~\eqref{eq:Laplace} must be done on a real axis in order to ensure the reality of the Borel resummed
function.  We must therefore distinguish whether singularities of the Borel transform are present or not on this integration
axis: without singularities, Laplace transforms with neighboring axis define the same analytic function
which can therefore be continued to a larger domain.  If there are singularities on this natural integration axis,
the function to integrate is obtained by an averaging procedure. This function is no longer
everywhere analytic, so that the integration axis cannot be shifted any more.

\section{Elements of Accelero-summation} 

This section intends to be a self-contained introduction to the basic concepts of accelero-summation.
We refer the reader to \cite{Ec92} for a more detailed description.

As is usual, the Borel transformation is defined as a map between two rings of formal series:
\begin{eqnarray} \label{eq:Borel_tsfo}
             \mathcal{B}: z^{-1}\mathbb{C}[[z^{-1}]] & \longrightarrow & \mathbb{C}[[\xi]] \\
\tilde{f}(z) = \frac{1}{z}\sum_{n=0}^{+\infty}c_n\frac{1}{z^n} & \longrightarrow & \widehat{f}(\xi) = \sum_{n=0}^{+\infty}\frac{c_n}{n!}\xi^n. \nonumber
\end{eqnarray}
We are interested in the cases where $\widehat f$ is a resurgent function\footnote{This means that the defining power series is convergent
in some neighbourhood of 0 and that the holomorphic function thus defined can be continued to the whole complex plane apart from a
discrete set of singularities.}. In cases where $\widehat f$ has a subexponential 
behaviour along an axis going from the origin to infinity, the Borel transform \eqref{eq:Borel_tsfo}
has an inverse, the Laplace transform, which defines an analytic function with an asymptotic behavior
at infinity given by the starting formal series. This whole procedure is 
called Borel resummation.

Acceleration allows to generalise this procedure to more complex situations, where the behaviour at
infinity of the Borel transform does not allow for a Laplace transform. Let $F:\C\mapsto\C$ be a function, which will be called the
acceleratrix, such that
\begin{itemize}
 \item $\overline{F(y)}=F(\bar y)$,
 \item $\lim_{y\to\infty}F(y)=+\infty$,
 \item $\lim_{y\to\infty}\frac{y}{F(y)}=+\infty$.
\end{itemize}
Let us specify that this last condition makes us deal with the so-called ``strong acceleration'' of
\cite{Ec92}, other forms exist but we will not need them.  Acceleration allows to have a Borel transform with respect to the variable
\(y\) starting from the Borel transform with respect to the variable \(z=F(y)\).

Acceleration is performed by mapping the function \(\widehat f(\xi)\) to a germ $\widehat
f_{\acc}(\zeta)$ by
\begin{equation}  \label{def_acc}
 \widehat f_{\acc}(\zeta):=\int_0^{+\infty}C_F(\zeta,\xi)\widehat f(\xi)d\xi
\end{equation}
The acceleration kernel is the Borel transform (with respect to the variable $y$) of the function $\exp(-\xi F(y))$ and can be 
obtained through an inverse Laplace transform: 
\begin{equation*}
 C_F(\zeta,\xi):=\frac{1}{2i\pi}\int_{c-i\infty}^{c+i\infty}e^{-\xi F(y)+\zeta y}dy ,
\end{equation*}
where \(c\) can be taken as any positive constant.

Similarly to the case of \(\widehat f\) which was a priori only defined in a neighbourhood of 0, the preceding integral only defines $\widehat f_{\acc}(\zeta)$ in the
vicinity of the origin since the integral in~\eqref{def_acc} is only convergent for small enough \(\zeta\) and we follow the same pattern. We analytically continue it, look for
singularities of this analytical continuation which are controlled by a new set of alien derivatives
and finally perform 
a Laplace (or median Laplace) transform on it to obtain the resummed function\footnote{In principle,
other round of accelerations could be necessary before the final Laplace transform, but once again, we
do not aim at describing the most general procedure.}.  The resummed function is then given by the following integral
\begin{equation} \label{eq:Laplace}
 f^{\res}(y) = \int\widehat f_{\acc}(\xi)e^{-y \xi}d\xi.
\end{equation}

All accelerations have in common that they transform
convolution products in convolution products, ensuring that the whole procedure will give a sum which
satisfies the same equations as the formal series we start with.  Strong acceleration allows to deal
with the case where the analytical continuation of the Borel transform is growing too fast to allow for
a Laplace transform.  Other forms of acceleration exist which allow to deal with functions that have other kind of defects, but we will
not try to present them here, since we do not need them.

In case of singularities on the preferred integration axis, the use of a convolution preserving average,
like in median resummation allows to obtain a real sum and has been successfully applied to various 
physical problems.

All transformations  that we use are algebra morphisms, either between convolution products or from the
pointwise product to the convolution product or reciprocally.  The chain of transformations begins by a Borel transformation and end by a
Laplace transformation with any number of averages and accelerations in the sequence, so that we can establish that the result of the
procedure satisfies exactly the
original equations, strongly limiting the ambiguities in the resummation procedure.

\section{Application to Quantum Field Theory}

In quantum field theory, the perturbation series is usually given in terms of a fine structure constant $a$, proportional to 
$g^2$. In order to be coherent with the notations of section 2, we will perform our Borel transformation in the variable $z=g^{-2}$.

Let us assume that the Borel transform is such that we need an acceleration of the form 
\begin{equation} \label{eq:acceleration}
 z = F(y) = \frac{1}{\sigma}\log(y).
\end{equation}
After resummation, one obtains a function $y\to\tilde G^{\res}(y)$ analytic in a half-plane $H_R:=\{y\in\C:\Re(y)\geq R\}$ for some $R>0$. Writing
$\tilde G^{\res}(y)=\tilde G^{\res}(F^{-1}(z))=:G^{\res}(z)$, we obtain that $z\to G^{\res}(z)$ is analytic in the domain $F(H_R)$.

Using the principal branch of the logarithm $\log(x+it) = \log(|x+it|)+i\arctan(t/x)$ (which was already used in Section 1) we see that the image under $F$
of a vertical line of real part $x\geq R$ is the curve parameterized by \(s=\arctan(t/x)\)
\begin{equation}
 S_x:=\left\{\left.\frac{1}{\sigma}\Bigl(\log(x)-\log(\cos(s))+is\Bigr)\right|s\in\left]-\frac{\pi}{2},\frac{\pi}{2}\right[\right\}.
\end{equation}
Since \(\log(\cos(s))\) goes logarithmically to minus infinity when \(s\) approaches \(\pm\pi/2\), 
the analyticity domain of $z\to G^{\res}(z)$ {is} well approximated by the open rectangle
\begin{equation}
 F(H_R) = \left\{z\in\C\left|\Re(z)\geq\frac{1}{\sigma}\log(R)\wedge\Im(z)\in\left]-\frac{\pi}{2\sigma},\frac{\pi}{2\sigma}\right[\right.\right\}.
\end{equation}
To relate this result to a more familiar setting, we have to map it back into the $g^2$ plane. 

The three lines approximating the boundary of \(F(H_R)\) will be converted to circles including the
origin, as was recalled in section~1.  First, \(F(H_R)\) is beyond the line with \(\Re(z)=
\log(R)/\sigma\), so its transformation will be inside the circle with center \(2\sigma/\log(R)\) and
radius the same expression.  The lines with \(\Im(z) = \pm \pi/(2\sigma)\) will likewise be transformed
in circles with centers \( \pm i \sigma/\pi \) and radius \(\sigma/\pi\), but this time, the image of
\(F(H_R)\) will be outside these circles.  All in all, we obtain in the usual \(g^2\) plane a domain
squeezed between the two tangent circles centered at \(\pm i \sigma /\pi\) near the origin, limited by
the circle centered at  \(2\sigma/\log(R)\).  We must not forget that the three circle limits are but
approximations, since the real boundary must be smooth, since it is the image of line by a holomorphic
map.

In conclusion, we have found that the accelero-summation gives the horn-shaped domain predicted by 't Hooft argument. 

\section{Acceleration and asymptotics of the Borel transform}

In the acceleration process, the essential parameter is the constant $\sigma$ appearing in the
acceleratrix $F(y) = \frac{1}{\sigma}\log(y)$.

For the analyticity domain obtained by accelero-summation not to be larger than the one given by 
't Hooft's argument, we must have $\sigma/\pi\geq K_0$, with $K_0$ given by \eqref{eq:limiting_circles}. We obtain
\begin{equation*}
 \sigma \geq \frac 1 {2 \beta_0}.
\end{equation*}

We can now put a bound on the asymptotic behaviour of the (non accelerated) Borel transformed two points function $\widehat G$. From the chosen accelerating function
$F(y) = \frac{1}{\sigma}\log(y)$ we can compute the acceleration kernel:
\begin{equation*}
 C_F(\xi,\zeta) = \frac{1}{2i\pi}\int_{c-i\infty}^{c+i\infty}z^{-\xi/\sigma}e^{\zeta z}dz =
 \frac{-1}{2i\pi}(\zeta)^{\xi/\sigma-1}\int_{-\zeta c-i\infty}^{-\zeta c+i\infty}(-y)^{-\xi/\sigma}e^{-y}dy               
\end{equation*}
where we have performed the substitution $y=-z\zeta$ of the integration variable and changed the
orientation of the integration contour. Since 
the integrand has no singularities outside the positive real line and goes to the zero sufficiently
rapidly on large quarter circles with positive real parts, we can deform the integration contour to one starting from 
$+\infty-i\epsilon$, going around the origin and ending in $+\infty+i\epsilon$. We recognize the Hankel contour of the reciprocal 
Gamma function $x\mapsto1/\Gamma(x)$. Hence we obtain
\begin{equation}
 C_F(\xi,\zeta) = \frac{(\zeta)^{\xi/\sigma-1}}{\Gamma(\xi/\sigma)}
\end{equation}
which is Equation (2.3.19) of \cite{Ec92}.  It is easy to verify that the Laplace transform with
respect to \(\zeta\) of \(C_F\) would give \(y^{-\xi/\sigma}\) which is exactly \(\exp[-\xi
F(y)]\). In the case where the integration on \(\xi\) and~\(\zeta\) is convergent on the whole domain $\R_+\times\R_+$, this ensures that we
recover the Laplace transform of the first Borel transform.  But this is certainly not the interesting
case.

The asymptotic behavior of \(C_F\) is easy to obtain from the Stirling approximation of the
\(\Gamma\)-function.  We have
\begin{equation}
\label{C_Fas}
	 C_F(\xi,\zeta) \mathop{\sim}_{\xi\to\infty} e^{\xi/\sigma\bigl( 1 - \log(\xi) +
	 \log(\zeta)\bigr)}
\end{equation}
up to less than exponential terms.  The good value of \(\sigma\) is then the one that makes the defining
integral for the accelerated germs absolutely convergent for some but not all values of \(\zeta\), that
is if 
\begin{equation} \label{bound_asymp}
  \log \bigl(\hat G(\xi) \bigr) \mathop{\sim}_{\xi\to\infty}\frac{1}{\sigma} \xi \log(\xi)
\end{equation}

\section*{Conclusion}

We have made explicit the statements of \cite{BeCl16}, namely that accelero-summation can
produce the analyticity domain of an 
asymptotically free quantum field theory predicted by 't Hooft's argument. To put this into context, we have also provided a rehearsal of 
't Hooft's argument and a short introduction to the essential idea of accelero-summation procedure.

Finally, we have been able to put a bound on the expected asymptotic behaviour of the Borel transformed two-point function for which
the accelero-summation procedure would give the right analyticity domain and it seems plausible that this bound is saturated.

Let us also observe that we have used here a quite particular version of accelero-summation. If the bound 
\eqref{bound_asymp} were not fulfilled, other forms of acceleration could be used.  In
particular,  if the Borel transform is equivalent to the exponential of a power of \(\xi\), acceleration
corresponding to \(F(y)=y^\alpha\) for \(\alpha\) a well-chosen positive real constant smaller than 1
allow to proceed.  However one easily checks that these accelerations allow to define the function in a sector of finite opening around
the origin. This 
suggests that the acceleration \eqref{eq:acceleration} is the right one in our case.

To conclude, let us emphasize that median resummation, while not discussed here, is totally
compatible with the accelero-summation procedure.
Indeed, if a non-perturbative mass can be generated as for example in \cite{BeCl16}, the Borel
transform must have singularities on the positive real axis, implying that real solutions can only be
obtained by using a suitable average of the analytic continuations. 
This suggests that all the 
tools needed to tackle the problem of non-perturbative mass generation in physically relevant quantum field theories are now at hand. It 
is toward this task that our future investigation shall lead.

\bibliographystyle{unsrturl}
\bibliography{renorm}

\end{document}